\documentclass{PoS}

\title{Double-Chooz Neutrino Experiment}

\ShortTitle{Double-Chooz}

\author{\speaker{Carmen Palomares}\thanks{On behalf of the Double-Chooz Collaboration}\\
        CIEMAT (Spain)\\
        E-mail: \email{mc.palomares@ciemat.es}}

\abstract{The Double Chooz experiment will use the electron antineutrinos produced by the Chooz nuclear power station to search for a non-vanishing value of the $\theta_{13}$ neutrino mixing angle.
Double Chooz will be the first of a new generation of neutrino experiments using identical detectors at different distances from the neutrino source to reduce the systemtic errors due to the uncertainties on the neutrino flux and to the detector acceptance. The far detector is expected to be operative by the beginning of 2010. Installation of the near detector will occur in 2010.}

\FullConference{European Physical Society Europhysics Conference on High Energy Physics,
EPS-HEP 2009,\\
		 July 16 - 22 2009\\
		 Krakow, Poland}

\def\t13{$\theta_{13}$}
\def\ane{$\bar{\nu}_{e}$~}

\begin{document}

\section{Physics Motivation}
Experiments with solar, atmospheric, reactor and beam neutrinos have provided compelling evidence for the existence of neutrino oscillations driven by non-zero neutrino masses and neutrino mixing~\cite{oscillation}. The mixing in the leptonic sector can be parametrised in terms of 3 mixing angles ($\theta_{12}$, $\theta_{23}$, $\theta_{13}$) and the CP violating phase $\delta_{CP}$. $\theta_{12}$ and $\theta_{23}$ have been measured to be large (almost maximal) while the third mixing angle, $\theta_{13}$, is only known to be small and $\delta_{CP}$ is totally unknown. 

The best limit of $\theta_{13}$ comes from the CHOOZ experiment~\cite{chooz}. The current bound from a global analysis of existing data~\cite{schwetz} is $\sin^{2}{2\theta_{13}}<0.13$ at 90\% C.L. Genuine three flavor oscillation effects occur only for a finite value of $\theta_{13}$. In addition, leptonic CP violation is also a three flavor effect, but it can only be tested by forthcoming experiments if $\theta_{13}$ is finite. Therefore, establishing a finite value of this mixing angle is one of the next milestones in neutrino physics.

Reactor neutrino experiments are able to provide a clean measurement of $\theta_{13}$ as they do not suffer, unlike accelerator experiments, from degeneracies and correlations between different oscillation parameters.

\section{Nuclear reactor experiments}
The nuclear reactor experiments look for the disappearance of electron antineutrinos produced in reactor cores with an average energy of 3 MeV. In order to measure $\theta_{13}$, a distance to the detector of about 1 km is chosen to maximize the disappearance probability to tau neutrino and to do negligible the oscillation to another flavor. This short baseline also prevents reactor experiment measurements to be affected by matter effects.

The \ane are detected via the inverse $\beta$-decay reaction: $\bar{\nu}_{e} + {\rm p}\rightarrow {\rm e}^{+} + {\rm n}$. The signature is a delayed coincidence between the prompt e$^{+}$ signal and the signal from the neutron capture. The positron energy is related to the \ane one by:  $E_{{\rm e}^{+}} = E_{\bar{\nu}_{e}}-(M_{\rm n}-M_{\rm p})$ and the $\gamma$-ray energy released after the n-capture goes from 2 MeV (capture on H) to 8 MeV (capture on Gd). 

This signal may be mimicked by background events which can be divided into two classes: accidental and correlated. The single events due to radioactivity from detector materials and surrounding rock, with a rate on the order of few Hz, is the responsible of the accidental background. The random coincidence between these events and a cosmogenic neutron would be drastically reduced by a careful selection of the materials used to build the detector. On the other hand, the two main contributions to the correlated background are: fast external neutrons and unstable isotopes, such as $^{9}$Li and $^{8}$He, with a $\beta$-neutron cascade. Fast neutrons may enter the detector and create recoil protons, faking the e$^{+}$ signal, and be captured after thermalization.
Both types of events are coming from spallation processes of high energy muons. However, in the second case, the correlation with a muon signal 
is very difficult due to the relatively long-live of these isotopes (about 0.1 s).

\section{The Double Chooz concept}
A new experiment which is designed to look for non-zero values of \t13 needs to go beyond the previous systematic limitations. The dominant systematic error from the absolute measurement of the reactor neutrino flux will be removed by using an additional detector located at few hundred meters from the nuclear core to monitor the unoscillated neutrino flux. In addition, if the detectors are identical the uncertainty related to detector acceptance will be largely reduced. 

The Double Chooz experiment~\cite{dchooz} is located close to the twin reactor cores of the Chooz nuclear power station in the Ardennes (France). The far detector will be placed at 1050 m distance from the cores in the same laboratory used by the CHOOZ experiment. It provides a quickly prepared and well-shielded (300 m.w.e.) site with near-maximal oscillation effect. The second identical detector (near detector) will be installed at 410 m away from the nuclear cores under a hill of 115 m.w.e.  

The detector has been designed to double the fiducial volume of the previous CHOOZ detector and the data taking period will extend to 3-5 years, reducing the CHOOZ statistical error (2.8\%) in a factor 5. To achieve this, the liquid scintillator stability has been successfully tested over 3 years.

\begin{figure}
\begin{center}
\includegraphics[width=.45\textwidth]{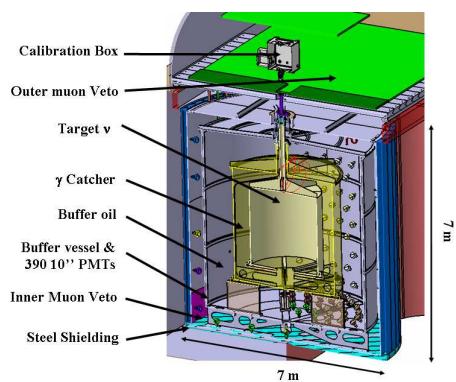}
\caption{The Double Chooz detector}
\label{fig:detector} 
\end{center}
\end{figure}

The Double Chooz detectors (see figure~\ref{fig:detector}) consist of concentric cylinders. 
The innermost volume is the neutrino target, a 10.3 m$^{3}$ acrylic cylinder, filled with 0.1\% Gd loaded liquid scintillator.
In addition, there is a 105 cm buffer of non scintillating oil to decrease the rate of single events from photomultiplier tubes (PMT) radioactivity. 
The detection system consists of 390 10 inches PMTs, providing about 13\% photocathode coverage.
The detector is encapsulated within an active cosmic-ray muon veto: a 50 cm thick region filled with liquid scintillator and instrumented with 78 8 inches PMTs.
Outside this vessel a 15 cm thick low activity steel shielding will protect the detector from the natural radioactivity of the rocks around the pit.
Finally, the upper part of the detector will be covered by plastic scintillator detector as an additional {\em outer muon veto}.

In Double Chooz, the error due to the uncertainty on the antineutrino flux will be replaced by the relative normalization between detectors. 
The goal of the experiment is to reduce this uncertainty to 0.6\%.
An effort is done for a precise measurement of the number of free protons inside the target volumes, since the antineutrino rate is proportional to it. 
Experimentally the target mass will be determined at 0.2\% and only one batch of liquid scintillator will be used to fill both detectors. The optimization of the detector design allows to simplify the analysis and to reduce the detection efficiency systematic error up to 0.5\%. 

The signal to background ratio will be kept above 100 for the near detector and above 20 for the far one, being the main contribution the correlated background. Under this condition, even a knowledge of the backgrounds within a factor two keeps the associated systematic error below the percent.

\begin{figure}
\begin{center}
\includegraphics[width=.5\textwidth]{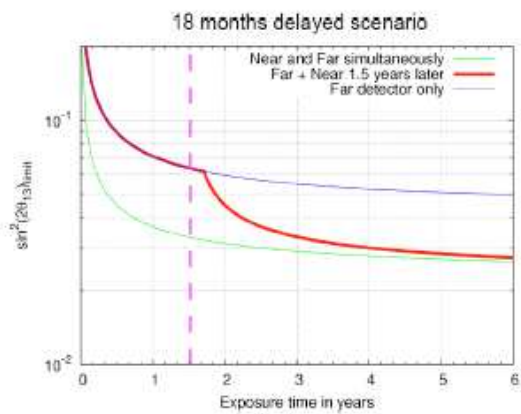}
\caption{Double Chooz expected sensitivity limit (90\% C.L.) to $\sin^{2}{2\theta_{13}}$ as a function of time.}
\label{fig:sens}
\end{center} 
\end{figure}

\section{Expected detector performance-sensitivity}
The data taking of Double Chooz will be divided in two phases. During the first one (starting at the beginning of 2010) only  the far detector will be present during 1.5 years. The expected systematic error is 2.5\%, where the main contribution (about 2\%) comes from reactor neutrino flux uncertainty. Figure~\ref{fig:sens} shows the improvement on the limit to 90\% C.L. on $\sin^{2}{2\theta_{13}}$ to be 0.06 with the far detector only, a factor 2 better than current limit. 
In the second phase both detectors will take data for 3 years, achieving a final sensitivity of $\sin^{2}{2\theta_{13}}<0.03$ in the case of non-oscillation.

\section{Conclusions}
The Double Chooz experiment is expected to start data taking at the beginning of 2010 to measure the angle \t13. After 3 years of operation of both detectors Double Chooz will be able to measure \t13 with 3$\sigma$ effect if  $\sin^{2}{2\theta_{13}}>0.05$. 
An intense R\&D work has been carried out by the Double Chooz collaboration to validate the robustness of the double detector concept.

\end{document}